\documentclass[12pt,showpacs,aps,prl,preprint]{revtex4}

\usepackage{amsmath}
\usepackage{mathrsfs}
\usepackage{bm}
\usepackage[pdftex]{hyperref}
\usepackage{graphicx}

\begin{document}
\begin{titlepage}
	
\title{Elliptic Cylindrical Invisibility Cloak, a Semianalytical Approach Using Mathieu Functions}
\author{E.~Cojocaru}
\affiliation{Department of Theoretical Physics, IFIN~HH,
Bucharest-Magurele MG-6, Romania}

\email{ecojocaru@theory.nipne.ro}

\date{today}

\begin{abstract}
An elliptic cylindrical wave expansion method by using Mathieu functions is developed to obtain the scattering field for a two-dimensional elliptic cylindrical invisibility cloak. The cloak material parameters are obtained from the spatial transformation approach. A near-ideal model of the invisibility cloak is set up to solve the boundary problem at the inner boundary in the cloak shell. The proposed design provides a more practical cloak geometry when compared to previous designs of elliptic cylindrical cloaks.
\end{abstract}


\maketitle
\end{titlepage}
Recently, achieving the invisibility of objects has received increased attention. Designs based on spatial coordinate transformations have been proposed \cite{re1,re2,re3,re4,re5,re6,re7,re8}. The inhomogeneous and anisotropic cloaks obtained from the spatial transformation bend the incoming waves around the cloaked region, so that the fields after emerging from the cloak are the same as if the incident waves had passed through the free space. The cloaking principle was demonstrated experimentally in the microwave regime \cite{re4}. Cloaks designs have been reported and investigated mostly for spherical and circular cylindrical geometries primarily due to the simplicity of analysis for structures that possess radial and axial symmetries. Recently, elliptic cylindrical cloaks have been reported based on non-orthogonal coordinate systems, the cloak being described by permittivity and permeability tensors with non-zero off-diagonal terms and verification being done numerically by full-wave finite-element simulations \cite{re9,re10}.

In this Letter, we will study the scattering for an elliptic cylindrical cloak by using a semianalytical approach that is based on Mathieu functions. By taking the advantage of the elliptic cylindrical geometry of the structure, we focus our analysis on the two-dimensional (2D) elliptic cylindrical cloak, because the wave equation can be simplified in comparison with the three-dimensional (3D) case. As in the circular cylindrical case \cite{re8}, we introduce a small perturbation into the ideal cloak to avoid extreme values (zero or infinity) of the material parameters at the cloak's inner surface.

First, let us look at the wave equation inside the elliptic cylindrical cloak. In terms of the rectangular coordinates $(x,y, z)$, the elliptic cylindrical coordinates $(u,v,z)$ are defined by the following relations \cite{re11,re12}
\begin{equation}
\label{eq:coord}
x = f \cosh u \,\cos v,\quad  y = f \sinh u\, \sin v,\quad  z = z ,
\end{equation}
with $0 \leq u < \infty$, $0 \leq v \leq 2\pi$, and $f$ the semifocal length of the ellipse. The contours of constant $u$ are confocal ellipses, and of constant $v$ are confocal hyperbolas. The $z$ axis coincides with the cylinder axis.
Similarly to the circular cylindrical cloak \cite{re1,re8}, a simple transformation $(u',v',z')$ with
\begin{equation}
\label{eq:transf}
u' = \frac{b-a}{b} \,u + a, \quad v'= v, \quad z'= z 
\end{equation}
can compress space from the elliptic cylinder region $0 < u < b$ into the elliptic shell $a < u' < b$, $u' = a$ being the inner confocal ellipse. Note that $u$, $u'$, $a$, and $b$ are dimensionless quantities. Let the position vector in the original system be written as $\vec{r} = \hat{x}x + \hat{y}y + \hat{z}z$ and in the transformed system as $\vec{r} \,^\prime  = \hat{x}x' + \hat{y}y' + \hat{z}z'$. Following the notations in \cite{re13}, the scale factors $Q_j (j=1,2,3)$ of the transformation [Eq.~\eqref{eq:transf}] are found to be
\begin{align}
&Q_1 \equiv \frac {\mid \partial \vec{r}/\partial u'\mid}
{\mid \partial \vec{r} \,^\prime /\partial u'\mid}=\frac{(\cosh \,2u-\cos \,2v')^{1/2}}{T(\cosh \,2u'-\cos \,2v')^{1/2}} ,  \nonumber  \\
&Q_2 \equiv \frac {\mid \partial \vec{r}/\partial v'\mid}
{\mid \partial \vec{r} \,^\prime /\partial v'\mid}=\frac{(\cosh \,2u-\cos \,2v')^{1/2}}{(\cosh \,2u'-\cos \,2v')^{1/2}} , \label{eq:qi} \\
&Q_3 \equiv \frac {\mid \partial \vec{r}/\partial z'\mid}
{\mid \partial \vec{r} \,^\prime /\partial z'\mid}= 1, \quad 
\textrm{where} \quad T=(b-a)/b.  \nonumber  
\end{align}
Finally, the permittivity and permeability tensor components for the cloak shell can be given as
\begin{equation}
\label{eq:tensor}
\frac{\epsilon_{u'}}{\epsilon_0} = \frac{\mu_{u'}}{\mu_0} = T, \quad
\frac{\epsilon_{v'}}{\epsilon_0} = \frac{\mu_{v'}}{\mu_0} = \frac{1}{T}, 
\quad
\frac{\epsilon_{z'}}{\epsilon_0} = \frac{\mu_{z'}}{\mu_0} = \frac{\cosh 2u -\cos 2v'}{T(\cosh 2u'-\cos 2v')} ,  
\end{equation}
where air is assumed for the ambient environment and the interior regions. In the following we consider the transverse-electric (TE) polarized electromagnetic field (i.e., the electrical field only exists in the $z'$ direction). One obtains the following general wave equation governing $E_{z'}$ field in the cloak's elliptic cylindrical coordinates
\begin{align}
&\frac{2}{f^2 \left(\cosh 2u'- \cos 2v' \right)} \Bigg\{ \frac{\partial}{\partial {u'}} \left( \frac{1}{\mu_{v'}}\frac{\partial E_{z'}}{\partial {u'}} \right) +
\frac{\partial}{\partial v'} \left( \frac{1}{\mu_{u'}} \frac{\partial E_{z'}}{\partial {v'}} 
\right) \Bigg\} \nonumber \\   
& + \epsilon_{z'} \left( \frac{1}{\epsilon_0 \mu_0}          
\frac{\partial^2E_{z'}}{\partial {z'}^2} - \frac{\partial^2E_{z'}}{\partial t^2} \right)= 0 .
\end{align}
An $exp(-i\omega t)$ time dependence is assumed, where $\omega$ is the circular frequency. If we substitute Eq.~\eqref{eq:tensor} for $\epsilon_{z'}$, $\mu_{u'}$, and $\mu_{v'}$ we find
\begin{equation}
\label{eq:weq}
T\frac{\partial^2E_{z'}}{\partial {u'}^2}+\frac{1}{T}\frac{\partial^2E_{z'}}{\partial {v'}^2}+\frac{f^2 {k_t}^2}{2T}\left(\cosh 2u - \cos 2v'\right)=0 ,
\end{equation}
where ${k_t}^2={k_0}^2-{k_z}^2$, ${k_0}^2=\omega^2 \epsilon_0 \mu_0$, and $k_z$ is the $z$-component of the wave vector (for 2D case, $k_z=0$). Equation~\eqref{eq:weq} can be solved by a separation of variables $E_{z'}=S(v')R(u')$ and the introduction of a separation constant $c$,
\begin{equation}
\label{eq:ang}
\Big\{ \frac{\mathrm{d}^2}{{\mathrm{d}v'}^2} 
+ \left( c - 2 q \cos 2v' \right) \Big\} S(v') = 0 ,
\end{equation}
\begin{equation}
\label{eq:rad}
\Bigg\{T^2 \frac{\mathrm{d}^2}{{\mathrm{d}u'}^2} 
- \left( c - 2 q \cosh 2 \frac{u'-a}{T} \right) \Bigg\} R(u') = 0 ,
\end{equation} 
where $q=f^2k_t^2/4$. Outside the cloak Eq.~\eqref{eq:ang} is the same, whereas Eq.~\eqref{eq:rad} is replaced by
\begin{equation}
\label{eq:radout}
\Big\{ \frac{\mathrm{d}^2}{{\mathrm{d}u'}^2} 
- \left( c - 2 q \cosh 2u' \right) \Big\} R(u') = 0 .
\end{equation}
Equation~\eqref{eq:ang} is known as the angular Mathieu equation \cite{re11,re12,re14}. The solution is denoted by $S_{p\,m}(q,v,n)$ where $p,m$ denote even $(e)$ and odd $(o)$, $n$ denotes the order. For simplicity, in the following the dependence of $q$ is skipped. Equation~\eqref{eq:radout} is known as the radial Mathieu equation. Similarly to the circular cylindrical coordinates where the radial solution is expressed in terms of Bessel functions $J_\nu(\cdot), ~Y_\nu(\cdot),~H_\nu^{(1)}(\cdot), ~\textrm{and} ~H_\nu^{(2)}(\cdot)$, the radial Mathieu equation has four kinds of solutions \cite{re11,re12,re14}: $J_{p\,m},~Y_{p\,m},~H_{p\,m\,1}, ~\textrm{and} ~H_{p\,m\,2}$, where $p,m = e,o$. Equation~\eqref{eq:rad} transforms into Eq.~\eqref{eq:radout} with the change of variable $u' \rightarrow \tfrac{u'-a}{T} $. 
In the following the primes of $u'$ and $v'$ coordinates are dropped for aesthetic reasons. 

Now, let consider a plane wave is incident on the cloak. In the 2D case, the incident field can be expanded in the $u,v$ coordinates with the following expression \cite{re11,re12}     
\begin{equation}
\label{eq:pw}
E_z^{in}=\sqrt{8\pi}\sum_{n}i^n J_{p\,m} (u,n) S_{p\,m}(v,n) S_{p\,m}(\phi ,n) / N_{p\,m}(n) ,
\end{equation}
where $\phi$ is the angle of incidence measured from the $x$ axis in the $(x,y)$ plane, and $N_{p\,m}$ is the normalization constant. Since $Y_{p\,m}$ has a singularity at $u=a$, in order to circumvent this, we introduce a small perturbation to the ideal cloak \cite{re8}. We consider the inner boundary located at $u=a+\delta$ where $\delta$ is a very small positive number. Now the electric field in each region can be given by                         
\begin{align}
\label{eq:field}
(u>b) \quad E_z & = \sqrt{8\pi}\sum_{n}i^n J_{p\,m}(u,n) S_{p\,m}(v,n) S_{p\,m}(\phi,n)/N_{p\,m}(n)  \nonumber  \\
    & + i^n \alpha_{p\,m}^{(sc)}(n) H_{p\,m\,1}(u,n) S_{p\,m}(v,n) S_{p\,m}(\phi,n)  \nonumber  \\
    (a + \delta < u < b) \quad E_z & = \sqrt{8\pi}\sum_{n}i^n \alpha_{p\,m}^{(1)}(n) J_{p\,m}(\tfrac{u-a}{T},n) S_{p\,m}(v,n) S_{p\,m}(\phi,n)     \\
     & + i^n \alpha_{p\,m}^{(2)}(n) H_{p\,m\,1}(\tfrac{u-a}{T},n) S_{p\,m}(v,n) S_{p\,m}(\phi,n)  \nonumber   \\
     (u < a + \delta) \quad E_z & =\sqrt{8\pi}\sum_{n}i^n \alpha_{p\,m}^{(3)}(n) J_{p\,m}(u,n) S_{p\,m}(v,n) S_{p\,m}(\phi,n)  \nonumber
\end{align}      
where $\alpha_{p\,m}^{(j)}(n)(j=1,2,3)$ are coefficients for the resulting field inside the cloak. The tangential electric and magnetic fields $E_z$ and $H_v$ (which can be obtained from $E_z$ \cite{re11,re12}) should be continuous across the interfaces at $u=a+\delta$ and $u=b$, and the orthogonality of $S_{p\,m}$ allows waves in each order $n$ and each combination $p,m=e,o$ to decouple. One obtains

\begin{align}
\label{eq:coeff}
&\alpha_{p\,m}^{(1)}(n)=\frac{1}{N_{p\,m}(n)} ,  \nonumber  \\
&\alpha_{p\,m}^{(2)}(n)=\alpha_{p\,m}^{(sc)}(n)=\frac{J_{p\,m}(\delta/T,n)}{N_{p\,m}(n)H_{p\,m\,1}(\delta/T,n)} \frac{\mathscr{J}_{p\,m}(a+\delta,n)-\beta_{p\,m}(n)\mathscr{J}_{p\,m}(\delta/T,n)}{\beta_{p\,m}(n)\mathscr{H}_{p\,m\,1}(\delta/T,n)-\mathscr{J}_{p\,m}(a+\delta,n)} ,  \\
&\alpha_{p\,m}^{(3)}(n)=\frac{\beta_{p\,m}(n)J_{p\,m}(\delta/T,n)}{N_{p\,m}(n)J_{p\,m}(a+\delta,n)} \frac{\mathscr{H}_{p\,m\,1}(\delta/T,n)-\mathscr{J}_{p\,m}(\delta/T,n)}{\beta_{p\,m}(n)\mathscr{H}_{p\,m\,1}(\delta/T,n)-\mathscr{J}_{p\,m}(a+\delta,n)} . \nonumber
\end{align} 
In these expressions we introduced the log-derivative function $\mathscr{F}=F\,^\prime \!/F$, $F$ being $J_{p\,m}$ or $H_{p\,m\,1}$, the prime denoting the derivative with respect to $u$, and
\begin{equation}
\label{eq:beta}
\beta_{p\,m}(n)=\frac{\int_0^{2\pi}[S_{p\,m}(v,n)]^2/[\sinh^2 (a+\delta) + \sin^2 v]^{1/2} \,\mathrm{d}v}
{\int_0^{2\pi}[S_{p\,m}(v,n)]^2/[\sinh^2 (\delta/T) + \sin^2 v]^{1/2} \,\mathrm{d}v} ,
\end{equation}
where the integration is performed numerically. Note that if $\delta \rightarrow 0$, then $\beta_{p\,m}(n)\rightarrow 0$ when $\phi=0$. Now, keeping almost the same notations like in \cite{re8}, we obtain in the case of 2D circular cylindrical cloak for any order $l$, 
\begin{align}
\label{eq:circ}
&\alpha_l^{(1)}=\alpha_l^{(in)} , \nonumber  \\
&\alpha_l^{(2)}=\alpha_l^{(sc)}=\alpha_l^{(in)}\frac{J_l(k_0\delta/T)}{H_l^{(1)}(k_0\delta/T)} \frac{\mathscr{J}_l[k_0(a+\delta)]-\beta \mathscr{J}_l(k_0\delta/T)}{\beta \mathscr{H}_l^{(1)}(k_0 \delta/T)-\mathscr{J}_l[k_0(a+\delta)]} ,  \\
&\alpha_l^{(3)}=\alpha_l^{(in)}\frac {\beta J_l(k_0 \delta /T)} {J_l[k_0(a+\delta)]} \frac{\mathscr{H}_l^{(1)}(k_0\delta/T)-\mathscr{J}_l(k_0\delta/T)}{\beta\mathscr{H}_l^{(1)}(k_0\delta/T)-\mathscr{J}_l[k_0(a+\delta)]} , \nonumber
\end{align}
where $\alpha_l^{(in)}$ is the coefficient of the incident plane wave, $\beta=\delta/[T(a+\delta)]$, and the log-derivative function $\mathscr{F}=F\,^\prime \!/F$ now refers to the Bessel function $J_l$ and Hankel function $H_l^{(1)}$, the prime denoting the derivative with respect to the argument. One can see the perfect similarity of the closed-form relations ~\eqref{eq:coeff} and ~\eqref{eq:circ} for elliptic and circular 2D cylindrical cloaks.

\begin{figure}[h]
\centerline{\includegraphics[height=10cm]{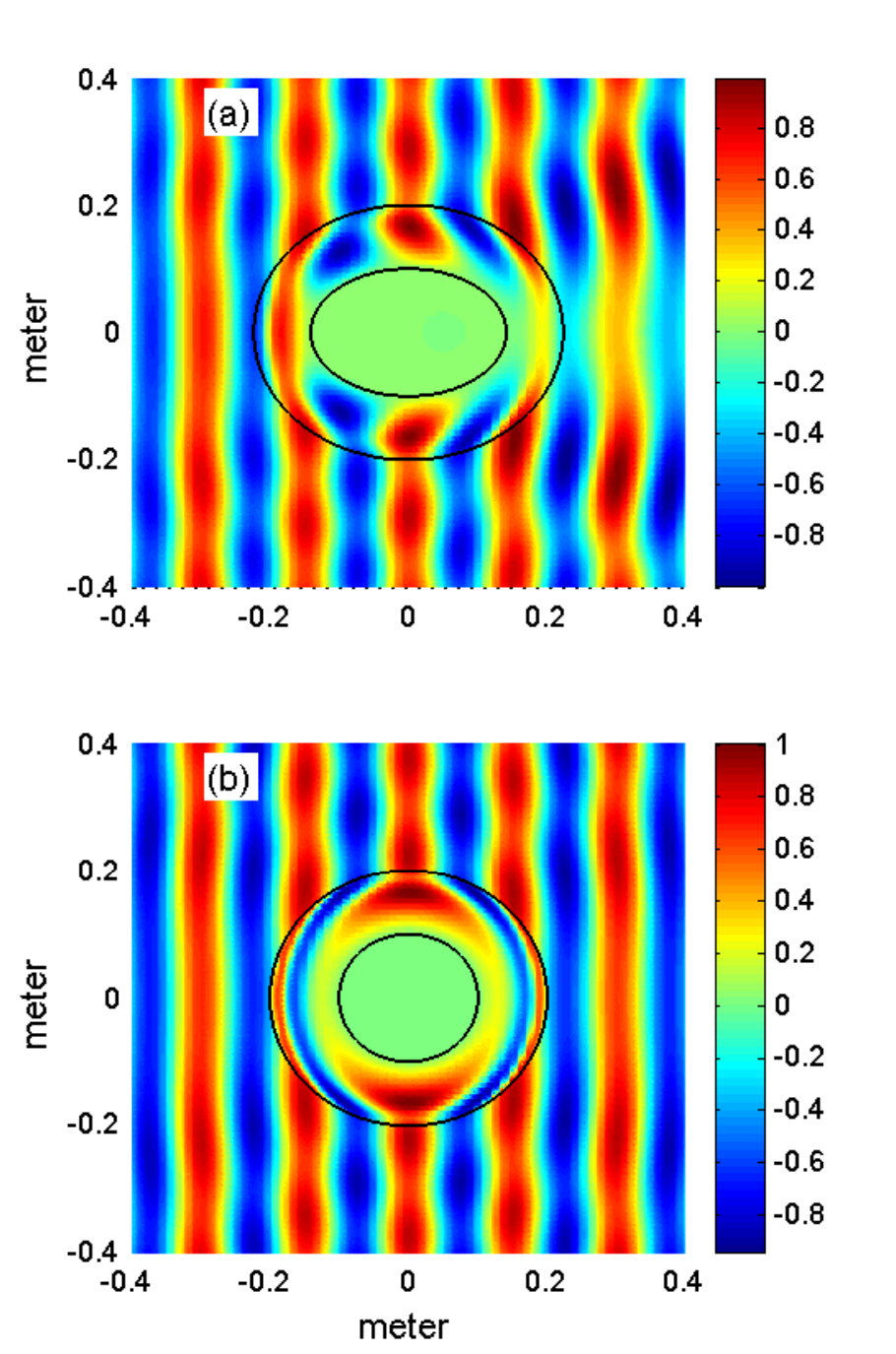}}
\caption{\label{fig:fone}Snapshots of the resulting electric field distribution in the vecinity of the cloaked object. The semiaxes on $(x,y)$ directions of the outer and inner ellipses are (0.2236m, 0.2m) and (0.1414m, 0.1m), respectively in (a), and (0.2002m, 0.2m) and (0.1005m, 0.1m) in (b). The incident light wavelength is 0.15m.}
\end{figure}

As an example, we consider $a$=0.8814, $b$=1.4436, $f$=0.1m, and $\delta=10^{-5}$. The semiaxes of the elliptic cylinder at $u=b$ are $d_x$=0.2236m, $d_y$=0.2m, and at $u=a$, $d_x$=0.1414m, $d_y$=0.1m. The wavelength of the incident plane wave in vacuum is $\lambda$=0.15m. We found that $n=0\div7$ is sufficient for convergence of the calculated scattering field \cite{re14}. Figure~\ref{fig:fone}(a) shows the snapshot of the resulting electric-field distribution (i.e., the real part of $E_z$ normalized against the maximum absolute value) in the vecinity of the cloaked object. The scatterer is illuminated from the $-\hat{x}$ direction. The cloak's two confocal elliptic boundaries (inner and outer) are shown as black contours in the figure. The electric-field distribution clearly demonstrates the cloaking effect of the cloak shell to the incident plane wave. The maximum absolute value of the normalized field inside the cloaked object is $7.5 \times 10^{-3}$. There is still a little bit of the field perturbation on the back side of the cloaking shell. This perturbation is diminished at smaller values of the semifocal distance $f$, that is, when the elliptic cylinder approaches a circular one. An example is shown in Fig.~\ref{fig:fone}(b), where $a$=2.9982, $b$=3.6895, and $f$=0.01m. Now, the semiaxes of the elliptic cylinder at $u=b$ are $d_x$=0.2002m, $d_y$=0.2m, and at $u=a$, $d_x$=0.1005m, $d_y$=0.1m. The maximum absolute value of the normalized field inside the cloaked object is $1.5 \times 10^{-4}$.

In conclusion, a semianalytical wave expansion method using Mathieu functions was applied to study the electromagnetic scattering properties of a 2D elliptic cylindrical invisibility cloak. A small perturbation into the ideal cloak was set up to solve the boundary problem at the inner surface of the cloak shell. Closed-form relations for the expansion coefficients of the resulting field inside the cloak were given which are perfect similar to those obtained for the circular cylindrical cloaks. Besides the theoretical interest for the variable separation problem solved by using Mathieu functions in case of elliptic cylindrical invisibility cloak shells, the results could be also useful for designing this type of invisibility cloak due to the very simple permittivity and permeability tensors, in spite of the non-perfect invisibility of the cloaks having elongated elliptic shapes.

\newpage

\end{document}